# Three-dimensional localization of spins in diamond using $^{12}$C implantation

Kenichi Ohno,[1] F. Joseph Heremans,[1,2] Charles F. de las Casas,[1,2] Bryan. A. Myers,[1] Benjamín J. Alemán,[1*] Ania C. Bleszynski Jayich,[1] and David D. Awschalom[1,2,†]

[1]*Center for Spintronics and Quantum Computation, University of California, Santa Barbara, California 93106, USA*

[2]*Institute for Molecular Engineering, University of Chicago, Chicago, Illinois, 60637, USA*

[*] Present address: Department of Physics, University of Oregon, Eugene, OR 97403, USA.

[†] To whom correspondence should be address. E-mail: awsch@uchicago.edu

**Abstract**

We demonstrate three-dimensional localization of a single nitrogen-vacancy (NV) center in diamond by combining nitrogen doping during growth with a post-growth $^{12}$C implantation technique that facilitates vacancy formation in the crystal. We show that the NV density can be controlled by the implantation dose without necessitating increase of the nitrogen incorporation. By implanting a large $^{12}$C dose through nanoscale apertures, we can localize an individual NV center within a volume of (~180 nm)$^3$ at a deterministic position while reproducibly preserving a coherence time ($T_2$) > 300 μs. Our approach enables integration of NV centers into diamond nanostructures to realize scalable spin-sensing devices as well as coherent spin coupling mediated by photons and phonons.

The single spin associated with the nitrogen-vacancy (NV) center in diamond provides an advantageous platform for spin-based quantum science and technology. The NV center's long spin coherence time ($T_2$) at room temperature[1] and the inherent atomic scale of the defect enable nanometer-scale magnetometry applications including single electron spin imaging[2,3] and sensing of ensemble nuclear spins external to the diamond.[4-6] In addition, the optical addressability of NV center spins provides a spin-light interface for quantum communications[7,8] leading to schemes of photon-mediated spin coupling.[9,10] These properties of the NV center can be further exploited through nanoscale engineering of the diamond. For example, fabricating the diamond into scanning probe devices for sensing,[11] as well as plasmonic cavities[12] and photonic crystals[13,14] can greatly enhance spatial resolution and photon collection efficiency. Shrinking mechanical structures containing NV centers to the nanoscale could also enable coherent strain-mediated spin interactions.[15]

A critical challenge to enabling scalable creation of NV-integrated nanostructures is to maintain spin coherence of an NV center localized at a deterministic position with nanoscale precision. Our approach utilizes a nitrogen delta-doping diamond growth technique which forms NV centers with consistently long spin coherence times,[16] while constraining the nitrogen and thus NV centers in the depth direction. We then use a post-growth vacancy creation process to control the lateral positioning of NV centers for full three-dimensional (3D) localization. In this letter, we demonstrate a method of vacancy engineering that allows for the tunability and localization of the NV density.

The previous approach to vacancy creation in delta-doped films was via high-energy (2 MeV) electron irradiation and subsequent annealing.[16] However, a drawback to this method is that vacancies are formed throughout the entire grown film and substrate. These non-localized vacancies result in a low density of doped NV centers on the order of $10^{12}$-$10^{13}$ cm$^{-3}$, which limits device integration of these engineered NV centers. Moreover, the significant amount of nitrogen in the substrate prohibits simply increasing the irradiation dose to increase doped NV density, as it would yield more background NV centers contained within the substrate. Since substitutional nitrogen (P1 centers) are a major source of spin decoherence for NV centers,[17] the NV density must be increased without increasing the number of nitrogen atoms incorporated within the diamond crystal. This requirement necessitates vacancy engineering that aims to enhance the conversion efficiency of nitrogen atoms within the delta-doped layer to NV centers.

Here we report a $^{12}$C implantation technique that enhances NV density without increasing the amount of doped nitrogen. $^{12}$C ions are implanted shallower (<~ 8 nm) than the nitrogen delta-doped layer (51 nm below the surface) to create a localized layer of vacancies whose depth is controlled by the implantation energy. Subsequent annealing diffuses these vacancies into the nitrogen delta-doped layer to form NV centers. As $^{12}$C has no nuclear spin, NV coherence times will not be limited by a nuclear spin bath.[1, 18] We show that the doped NV density can be controlled primarily through the $^{12}$C implantation dose, and furthermore that this technique can be utilized with aperture implantation[19] to provide 3D localization of single NV centers in a volume of (~180 nm)$^3$ while still reproducibly preserving long spin coherence times $T_2$ > 300 μs.

All samples used in this study were grown using a previously demonstrated nitrogen delta-doping technique in a diamond plasma-enhanced chemical vapor deposition (PE-CVD) system.[16] We first grew a $^{12}$C buffer layer (51 or 103 nm thick) on top of a commercially available, electronic grade (100) diamond substrate from *Element 6*, followed by a 6 nm-thick $^{15}$N delta-doped layer and a $^{12}$C cap layer (51 nm). The thickness of each layer was estimated by our calibrated growth rate.[16] We doped with isotopically purified $^{15}$N$_2$ gas (> 98 %) in order to differentiate doped $^{15}$NV centers from $^{14}$NV centers in the substrate. After growth, the samples were implanted with $^{12}$C ions at an implantation energy of 2 or 7 keV. According to our SRIM calculations,[20] these implantation energies create vacancies shallower than 2.4 ± 2.2 or 8.0 ± 6.1 nm from the surface, respectively (see Figure 1A). Creating localized vacancies shallower than the nitrogen delta-doped layer is important to preserve spin coherence of doped NV centers, as the delta-doped layer is deep enough to be protected from crystal damage due to the ion impact. On the other hand, the vacancy layer is still close enough (10s of nm) to the nitrogen delta-doped layer that vacancies can diffuse during the subsequent annealing process in a H$_2$/Ar forming gas atmosphere, creating NV centers in the delta-doped layer. A $^{12}$C implantation dose was selected from a range of $10^9 - 10^{13}$ cm$^{-2}$ depending on the experiment. After the implantation and annealing processes, the samples were cleaned in perchloric mixture (HClO$_4$:HNO$_3$:H$_2$SO$_4$ = 1:1:1, at 190 $^\circ$C for 30 minutes) to remove a graphitic layer on the surface as well as to oxidize the diamond surface to promote the negatively charged NV$^-$ center.[21, 22] The sample preparation process and $^{12}$C implantation technique are summarized schematically in Fig. 1A. All measurements were performed using a home-built confocal microscopy setup with 532 nm laser excitation at ambient conditions.

First, we demonstrate that the NV centers in the nitrogen delta-doped layer are formed through a vacancy diffusion process. Figure 1B shows three confocal photoluminescence (PL) scans at the diamond surface as-implanted (left), after annealing at 800 °C for 30 minutes (center) and 1 hour (right). We used the same sample for this cumulative annealing study and imaged three randomly selected areas on the surface for each annealing step to count the number of formed NV centers. We used continuous-wave electron spin resonance (CW-ESR) measurements to identify the isotopic signature for doped $^{15}$NV centers (circled in red) and substrate $^{14}$NV centers (black).[16] The as-implanted surface showed no trace of NV centers, whereas after 30 minutes of annealing we observed that > 90 % of all investigated NV centers were doped NV centers. On the other hand, when we annealed for a total of 1 hour, we observed both doped (~70 % of measured NV centers) and substrate NV centers. These results are consistent with a picture in which the vacancies diffuse away from the implantation region, pass through the delta-doped layer activating doped NV centers, and ultimately diffuse deeper in the crystal to activate substrate NV centers. We note, however, that a quantitative understanding of the vacancy diffusion mechanism in diamond is necessary in order to form a procedure that exclusively activates doped NV centers. We also note that in other samples we were able to activate NV centers in a ~12 nm deep nitrogen delta-doped layer with an implantation energy of 2 keV, and their coherence times were suitable for external spin sensing.[23]

Next, we investigate how process parameters affect the doped NV density. We grew two samples of nominally identical structure with two different $^{15}N_2$ doping gas flow rates (10 sccm and 0.1 sccm). The samples were then implanted at 7 keV on three different quadrants with different $^{12}C$ doses ($10^9$, $10^{10}$, and $10^{11}$ cm$^{-2}$), followed by annealing at 850 °C for 30 minutes. The six figures shown in Fig. 2A are surface confocal scans of each quadrant. We identified doped NV centers (circled in red) through their CW-ESR hyperfine signature. We measured three random regions within each quadrant on both samples to build up statistics and calculated the doped NV density by computing the average number of doped NV centers found over the scanned area (10 μm x 10 μm) assuming a 6 nm thickness of the nitrogen doped layer. The numbers at the top left of each figure denote the doped NV density in units of $10^{13}$ cm$^{-3}$. For both $^{15}N_2$ doping gas flows, the doped NV density increased from ~ 0.1 to ~ 1 x $10^{13}$ cm$^{-3}$ when we increased the $^{12}C$ dose from $10^9$ to $10^{10}$ cm$^{-2}$, suggesting that the doped NV density can be controlled via $^{12}C$ implantation dose. According to secondary ion mass spectrometry (SIMS) measurements,[16] the nitrogen dopant concentration should be (0.8 ± 0.6) x $10^{16}$ cm$^{-3}$ for the 10 sccm $^{15}N_2$ doping gas flow, which is consistent with the nitrogen spin bath density extracted from NV coherence times in similar films.[24] We calculate the nitrogen to NV conversion efficiency ([NV]/[N]), where [NV] is the measured doped NV density and [N] is the estimated nitrogen dopant concentration; the results for the 10 sccm $^{15}N_2$ flow sample are 0.02 % and 0.16 % for $10^9$ and $10^{10}$ cm$^{-2}$ $^{12}C$ doses, respectively. These low conversion efficiencies suggest that NV center creation is vacancy-limited for these implantation dose regimes. For the $10^{11}$ cm$^{-2}$ $^{12}C$ implantation dose cases, the NV density was too high to isolate single doped NV centers, but the integrated photoluminescence (PL) over the scanned areas showed ~20 – 40 times (varied with measured positions) increase relative to the $10^{10}$ cm$^{-2}$ areas suggesting that doped NV density continues to increase with $^{12}$C implantation dose. On the other hand, the doped NV density did not show noticeable $^{15}$N$_2$ doping gas flow dependence in the $10^9$ and $10^{10}$ cm$^{-2}$ dose areas, even though it was varied over two orders of magnitude. These results challenge prior assumptions that nitrogen incorporation into the diamond crystal can be controlled by the doping gas flow,[16] rather suggesting that the nitrogen incorporation is limited in this regime and has already saturated.

We next investigated whether coherence times are affected by the change in vacancy creation density and whether the hypothesis of nitrogen incorporation saturation can be corroborated by the spin coherence. Figure 2B summarizes $T_2$ results, measured using a Hahn echo sequence, for NV centers created by the $^{12}$C implantation technique. We measured twenty individual NV centers, five for each of the four combinations of $^{12}$C implantation doses ($10^9$ and $10^{10}$ cm$^{-2}$) and $^{15}$N$_2$ doping gas flows (10 and 0.1 sccm). These $T_2$ times were reproducibly on the order of hundreds of microseconds. In particular, the longest $T_2$ of a single NV center, (799 ± 77) μs, was measured from the $10^9$ cm$^{-2}$ dose area of the 0.1 sccm flow sample and is encouragingly comparable to the results from electron irradiated samples.[16] Only two NV centers, both from $10^{10}$ cm$^{-2}$ dose area in the 0.1 sccm sample, showed $T_2 < 50$ μs. These outliers make the standard deviations of $T_2$ particularly large in the 0.1 sccm flow sample, and we hypothesize that they are likely caused by local fluctuations in the density of vacancy-related paramagnetic defects that form due to the implantation process and are not fully removed by our annealing procedure.[25]

The means and standard deviations (as error bars) of $T_2$ over each set of five NV centers are shown in the right panel of Fig. 2B. The average $T_2$ time increased from $182 \pm 18$ ($252 \pm 93$) μs to $235 \pm 38$ ($614 \pm 106$) μs for the 10 (0.1) sccm sample when the $^{12}$C dose is changed from $10^{10}$ to $10^9$ cm$^{-2}$. In order to investigate the effects of $^{12}$C dose and $^{15}$N$_2$ flow on $T_2$, we carried out a two-way analysis of variance of our complete $T_2$ data, which showed a statistically significant main effect for the $^{12}$C implantation dose ($p < 0.05$), while the main effect for $^{15}$N$_2$ flow and the interaction between the two main effects were not statistically significant ($p > 0.05$).[26] There may still be more complicated effects of vacancies and nitrogen affecting the $T_2$ times, which cannot be deterministically verified from the limited sample set.[26] However, if nitrogen were in fact incorporated proportionately to the nitrogen flow rate, we would expect $T_2$ times to be orders of magnitude different between NV centers in the 0.1 and 10 sccm flow samples.[17] The differences we see in the $T_2$ times are only within the same order of magnitude between samples and therefore the results are consistent with our conclusion from Fig. 2A that the nitrogen incorporation is weakly dependent on the doping gas flow.

We further utilize the $^{12}$C implantation vacancy creation technique to demonstrate 3D position control of NV centers within a nanoscale volume. To achieve a sufficiently high density of NV centers we increased the thickness of the nitrogen doped layer and increased the $^{12}$C implantation dose. We grew a sample containing a 51 nm-thick $^{12}$C buffer layer, followed by a 51 nm-thick $^{15}$N doped layer and a $^{12}$C cap layer (also 51 nm). After growth, we patterned an array of apertures of varying diameters (nominally from 50 to 450 nm) on a layer of spin-coated resist (PMMA 950) using an electron beam (EB) lithography technique that masks the rest of the surface area from implantation.[19] The sample was then $^{12}$C implanted with an increased dose of $10^{13}$ cm$^{-2}$ at an implantation energy of 7 keV, and subsequently annealed in H$_2$/Ar forming gas at 850 °C for 30 minutes. Finally, the sample was cleaned using the perchloric acid procedure previously described. The device process is summarized in Fig. 3A, and a scanning electron microscope (SEM) image of the patterned apertures is shown in the inset of Fig. 3B, which was taken from a separate sample processed simultaneously with the same procedure.

In Figure 3B, we show a surface confocal scan of the processed sample showing patterned PL from NV centers localized by implanting through apertures. The difference in integrated PL of seven columns (A - G, from left to right) reflects the difference in their aperture diameter (larger apertures on the left). In particular, the right two columns (columns F and G) show less PL intensity than other columns since they have smaller diameters, less than 140 nm as measured from the SEM image. We measured 90 apertures in each column to build up statistics in order to determine how many NV centers are localized per aperture. Figure 3D is a histogram showing occurrences of finding $n$ NV centers ($n$ = 0, 1, 2…10) per aperture (data shown in blue circles) in column G, which has measured diameter of 114 ± 21 nm. We observed 18 % of all measured apertures in this column showing integrated PL consistent with a single NV center. We used a maximum likelihood approach to fit a Poisson distribution (red circles) to these data, which gave a mean value of $\lambda$ = 0.88 NV centers per aperture.

In order to check the validity of our PL-based NV counting method, four of the NV centers showing PL consistent with single NV centers were each tested using photon anti-bunching measurements, including the one circled in red in Fig. 3B. The result on this NV center is shown in Fig. 3C, displaying the normalized second order correlation function without background correction at $t = 0$, $g^{(2)}(0) < 0.5$, which confirms we are observing luminescence from a single NV center. We measured the $T_2$ times of these four single NV centers. The result again showed a scattering of $T_2$ times between 60 and 530 μs, reflecting the non-uniform damaging nature of ion implantation. However, we measured an average $T_2$ of ~340 μs and two of these displayed long spin coherence times of $T_2$ ~500 μs, which are more than an order of magnitude longer than previously reported $T_2$ times for position-controlled NV centers formed using nitrogen ion implantation through apertures.[19] The result of the Hahn echo measurement for the NV center showing longest $T_2 = 530 \pm 46$ μs is shown in Fig. 3E.

From these results, we demonstrate that spin coherence can be retained in the three-dimensionally localized single NV centers. The largest uncertainty in estimation for the localized volume arises from vacancies diffusing not only in the depth direction, but also delocalizing in the in-plane direction. We estimate the vacancy diffusion length to be $153^{+123}_{-67}$ nm based on our observation that $88^{-18}_{+10}$ % of measured NV centers in the apertures were doped NV centers and assuming a one-dimensional diffusion model.[26, 27] If we assume that vacancies diffuse isotropically from the aperture, and use the mean value of the estimated vacancy diffusion length, we calculate the localized volume to be 5.7 x $10^6$ nm$^3$, or (179 nm)$^3$, and the corresponding NV conversion efficiency to be 1.9 %.[26]

The calculated conversion efficiency is increased only by an order of magnitude from the $10^{10}$ cm$^{-2}$ (Fig. 2A) to $10^{13}$ cm$^{-2}$ (Fig. 3B) dose cases. However, the conversion efficiency is affected by the localized volume, which sensitively depends on the estimation of vacancy diffusion length. Also, the incident angle of $^{12}$C ions was tilted 7 degrees, which might result in shadowing of some of the implanted ions by the nanoscale apertures. Confinement of the NV centers within nanopillars could enable a more precise estimation of the conversion efficiency.[26]

The estimated localized volume is comparable to the length scale of many types of NV-based nanostructures, encouraging integration of these $^{12}$C-implanted NV centers into devices. For example, photonic crystals fabricated from diamond can have mode volumes of a fraction of $(\lambda / n)^3$, where $\lambda = 637$ nm is the wavelength of the zero phonon line of an NV center, and $n = 2.4$ is the refractive index of diamond.[13, 14] This mode volume makes a monolithic NV-photonic crystal approach viable using our technique, which could lead to the integration of coherent NV centers into photonic network. As another example, proposals for coherent phonon-mediated spin coupling[15, 28] require the placement of NV centers close to the surface of nanoscale beams in order to maximize strain induced in the NV centers. Our 3D localization approach is particularly promising for this implementation because of the advantage of nanometer-scale depth control inherent to the nitrogen delta-doping technique.

In summary, we demonstrated a $^{12}$C implantation technique to control the NV center density in the nitrogen delta-doped layer of a CVD-grown diamond film. Localized vacancies are independently created at a depth shallower than the nitrogen doped layer by tuning the implantation energy and the vacancies subsequently diffuse to activate doped NV centers during the annealing process. These NV centers show long spin coherence times $T_2$ on par with those of NV centers created via electron irradiation. Furthermore, the NV center density can be controlled with the $^{12}$C implantation dose. Finally, by using a higher implantation dose through lithographically patterned apertures, we demonstrated that single NV centers can be localized in three-dimensions within a volume of $(\sim 180 \text{ nm})^3$ at a deterministic position while repeatedly retaining their spin coherence times $T_2 > 300$ μs. Our demonstration of the 3D localization of coherent NV centers is compatible with NV-based nanostructures, enabling scalable creation of spin sensing devices, photonic network of coherent spins and phonon-mediated spin entanglement. While most potential applications require single NV centers, this technique additionally affords the ability to create dense regions of NV centers, which could facilitate quantum registers based on magnetic dipole coupling of proximal spins. [29, 30]

This work was supported by DARPA and AFOSR. B.A.M. acknowledges support from a Department of Defense (NDSEG) fellowship. B.J.A. acknowledges support from the University of California President's Postdoctoral Fellowship. The authors thank Claire A. McLellan for experimental help, David J. Christle for statistical assistance and Audrius Alkauskas for useful theoretical discussion.

**Figure Captions**

Fig. 1. (Color) (a) Schematic of the $^{12}$C implantation technique process and sample structure. (b) Confocal photoluminescence (PL) scans of the diamond surface as-implanted (left), annealed for 30 minutes (center) and 1 hour (right) at 800 °C. Doped $^{15}$NV centers are circled in red, and substrate $^{14}$NV centers are circled black. PL spots that show CW-ESR indicative of multiple NV centers are not circled.

Fig. 2. (Color) (a) Surface confocal PL scans of six different parameter areas. The two rows correspond to different $^{15}$N$_2$ flow rates of 10 (top) and 0.1 sccm (bottom), and the three columns are for $^{12}$C doses of $10^9$ (left), $10^{10}$ (center), and $10^{11}$ cm$^{-2}$ (right). In the columns for the $10^9$ and $10^{10}$ cm$^{-2}$ doses, doped $^{15}$NV centers are circled in red and substrate $^{14}$NV centers are circled black. NV centers showing inconclusive results due to poor CW-ESR contrast are left uncircled. Single NV centers could not be isolated in the $10^{11}$ cm$^{-2}$ dose. The numbers in the top left of each figure denote the doped NV densities in units of $10^{13}$ cm$^{-3}$. (b) Hahn echo spin coherence times ($T_2$) of NV centers from $10^{10}$ (left) and $10^9$ cm$^{-2}$ (center) $^{12}$C dose areas. Each data point corresponds to the $T_2$ time of a single NV center. Uncertainties in $T_2$ are quoted at 95 % confidence. The average $T_2$ time of five NV centers from each process parameter are shown in the right panel. Blue circles are data for the 0.1 sccm $^{15}$N$_2$ flow sample and red squares are for the 10 sccm sample. The green belt represents $T_2$ results of electron irradiation diamond film samples from ref. 16.

Fig. 3. (Color) (a) Schematic of the device process to localize NV centers in three dimensions using an aperture array. (b) Surface confocal scan of the 3D localization sample, showing a grid of PL spots from NV centers localized in an array. One of the single NV centers is circled in red. (inset) SEM image of apertures of a separate sample processed simultaneously in the same recipe. (c) Photon autocorrelation function $g^{(2)}(\tau)$ as a function of delay time $\tau$ measured in the NV center circled in red in (b). (d) Histogram of finding $n$ NV centers ($n = 0, 1…10$) per aperture in a column in of smallest diameter ($114 \pm 21$ nm) in (b). The data are shown in blue circles and a fit by a Poisson distribution with a mean value of $\lambda = 0.88$ is shown in red circles. (d) Hahn echo decay data (blue points) and a fitting curve (red line) of one of the single NV centers showing longest $T_2 = 530 \pm 46$ µs. Uncertainties in $T_2$ are quoted at 95 % confidence.

Fig. 1

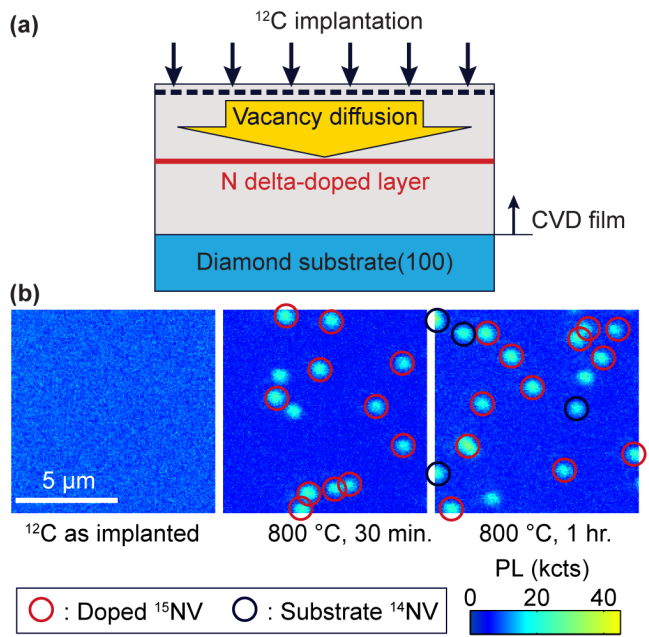

Fig. 2:

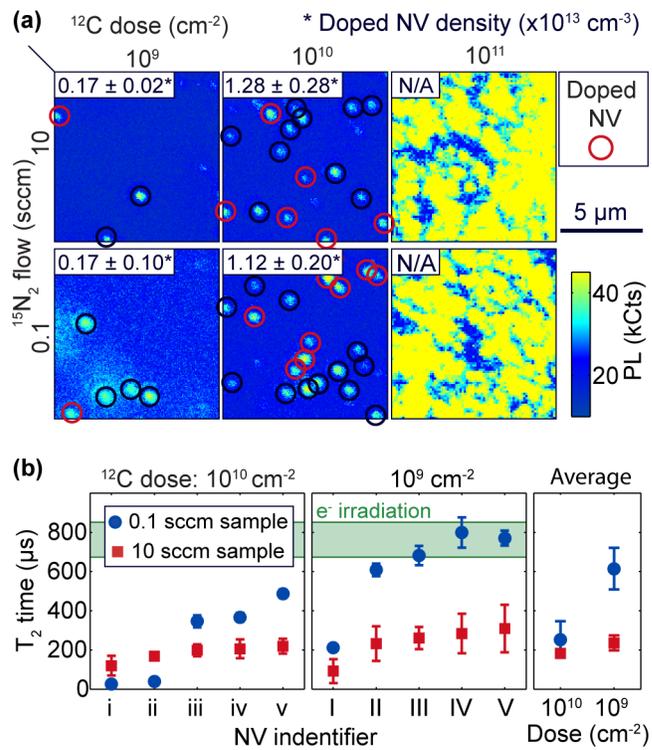

Fig. 3

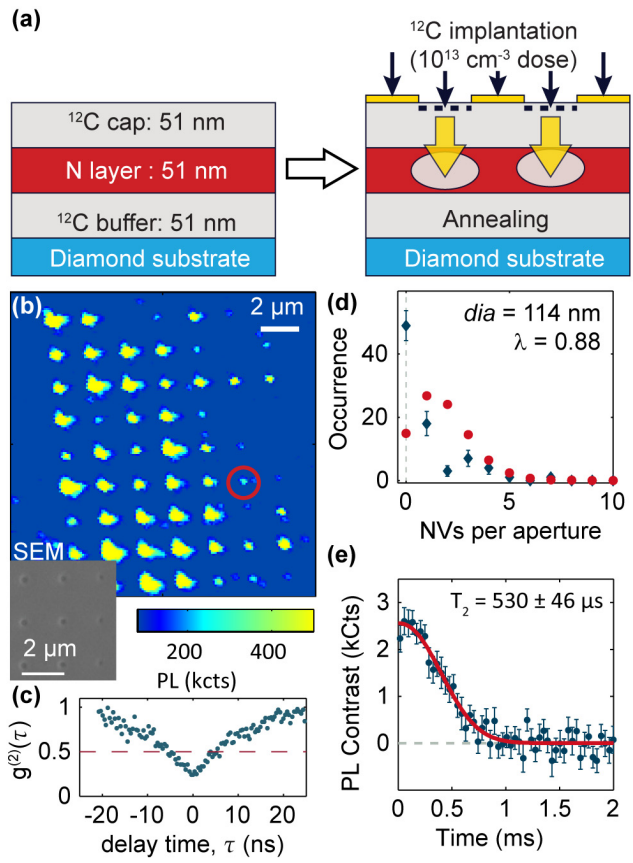